\documentclass{Odyssey2026}

\title{Multi-Axis Speech Similarity via Factor-Partitioned Embeddings}

\author[affiliation={1}]{Jim}{O'Regan}
\author[affiliation={1}]{Jens}{Edlund}
\interspeechcameraready

\affiliation{Department of Speech, Music \& Hearing}{KTH Royal Institute of Technology}{Sweden}
\email{joregan@kth.se, edlund@speech.kth.se}

\usepackage{comment}
\usepackage{soul}
\usepackage{hyperref}
\hypersetup{hidelinks}
\usepackage[utf8]{inputenc}
\begin{document}

\maketitle

\begin{abstract}
Speech encodes multiple simultaneous attributes--linguistic content, speaker identity, dialect, gender--that conventional single-vector embeddings conflate.

We present a factor-partitioned embedding framework that maps each utterance into a single vector whose subspaces correspond to distinct axes of variation.

A shared acoustic encoder feeds per-axis linear projection heads, each trained via distillation from a specialist teacher or a contrastive objective over shared-label pairs.

The resulting embeddings support attribute-conditioned retrieval: similarity is computed as a signed weighted sum over per-axis cosine scores, allowing retrieval that jointly considers what was said and how--or explicitly suppresses one attribute to surface another.

We evaluate on cross-corpus retrieval over corpora sharing the Harvard sentence prompts, demonstrating that signed axis weighting can suppress same-speaker bias and surface semantically matched utterances across recording conditions.

Code is available at:
\url{https://github.com/jimregan/spoken-sentence-transformers}
\end{abstract}

\section{Introduction}

Speech, in contrast with text, is inherently variable.
We never say the same thing the same way twice: different speakers say the same thing differently, and even the same speaker says the same thing differently at different times, due to a variety of factors such as mood, context, environment, and communicative intent.
Beyond linguistic content, speech reveals much about the speaker: more fundamentally, it can reveal their identity, sex, and even health \cite{nolan1995can}.
These attributes are often intertwined, and it can be difficult to disentangle them.

The prompts presented to speakers when creating text-to-speech datasets have traditionally been specifically selected to be phonetically balanced \cite{kominek2004cmu, veaux2013bvoicebank}. In multi-speaker datasets, there is often substantial overlap between the prompts read by different speakers.
This overlap presents an opportunity to evaluate the feasibility of attribute-conditioned similarity: given two utterances of the same sentence, an ideal representation should allow retrieval based not only on \textit{what} was said, but \textit{who} said it.

We introduce a factor-partitioned embedding framework for speech that learns a single embedding space, partitioned into subspaces corresponding to different attributes.
Given an utterance, the model produces a concatenated embedding vector whose subregions are specialised for different factors, such as semantic content and speaker identity.

Similarity in this framework is treated as a vector instead of a scalar: instead of computing a single similarity score between utterances, we compute per-axis similarities and combine them with a weighted sum. 
Retrieval can then be conditioned on different attributes: similarity can be computed jointly over multiple axes, or an attribute can be suppressed.
For example, assigning a negative weight to the speaker identity axis discourages same-speaker matches, and surfaces similar results from different speakers.

While speech embeddings have been studied extensively, most approaches produce a single similarity measure that conflates multiple attributes. To our knowledge, controllable multi-axis similarity search for speech has not previously been explored.

\section{Related work}

\subsection{Speech representation learning}

Self-supervised speech models, such as wav2vec 2.0 \cite{baevski2020wav2vec2}, HuBERT \cite{hsu2021hubert}, and WavLM \cite{chen2022wavlm} adapt the masked language model pre-training objective introduced by BERT \cite{devlin-etal-2019-bert} to learn general purpose representations for speech.

These models capture a mixture of linguistic, speaker, and prosodic information within a single representation; while this allows a single shared model to be fine-tuned on a variety of tasks \cite{yang21c_interspeech}, it also makes it difficult to control which attributes drive downstream similarity.

Speaker embeddings such as x-vectors \cite{snyder2018x} address this for speaker recognition specifically, but do not carry semantic content.

%Acoustic word embeddings and query-by-example retrieval
Acoustic word embeddings (AWEs) map fixed-length audio segments to vectors such that phonetically similar segments are nearby, enabling query-by-example spoken term detection without transcription \cite{kamper2016awe, he2017multiview, chung16_interspeech}. The retrieval evaluation paradigm--retrieve audio items matching a spoken query--motivates the cross-corpus setup used here, albeit at the utterance level rather than the word or segment level. AWE models optimise a single notion of acoustic similarity; our work extends this to multiple simultaneous axes, to allow retrieval to be steered toward semantic content, speaker identity, or other properties independently.

%Sentence transformers and multi-notion similarity
Sentence-BERT \cite{reimers2019sentence} adapts BERT-family models to produce fixed-size sentence embeddings suitable for semantic similarity via contrastive training\footnote{Oord, Aaron van den, Yazhe Li, and Oriol Vinyals. 2018. ``Representation Learning with Contrastive Predictive Coding.'' arXiv Preprint 1807.03748.}. The SentenceTransformers\footnote{\href{https://github.com/huggingface/sentence-transformers}{https://github.com/huggingface/sentence-transformers}} library generalises this pipeline to arbitrary encoder–pooler–projector stacks. We adapt this pipeline to speech by replacing the text encoder with an acoustic encoder (WavLM or similar) and introducing multiple attribute-specific projection heads.

Conditional Similarity Networks \cite{veit2017conditional} are the closest conceptual prior for our retrieval approach: they learn a single embedding space partitioned into subspaces corresponding to different kinds of similarity, and select or re-weight subspaces at query time via a mask or weight vector. Our work extends this idea to speech in three respects: (1) application to the speech modality, (2) signed weighting rather than binary masking, and (3) axis supervision via teacher distillation rather than triplet supervision.

The semantic axis is an instance of speech-to-text retrieval alignment: the speech encoder is trained to produce embeddings that agree with a text-side teacher, enabling content-based retrieval without relying on ASR at inference time. This pattern has been explored in SpeechDPR \cite{lin2024speechdpr} and SpeechRAG \cite{min2025speechrag}, which align speech encoders to frozen text dense retrievers. CLAP \cite{elizalde2023clap} takes a related approach at the audio level, training a joint audio–text embedding space via contrastive learning over audio clips paired with natural-language descriptions.

\subsection{Disentangled speech representations}

Our work has a degree of superficial similarity to work on disentangled representation learning; our goals, however, are more modest. Rather than enforcing factor independence within the encoder, we focus on supervised learning of multiple relational axes and train per-axis projection heads that induce distinct similarity geometries over a shared frozen acoustic representation space.

The earliest work on sequential disentanglement used factorised hierarchical variational autoencoders (FHVAEs) \cite{hsu2017unsupervised}, which separate sequence-level factors, such as speaker identity, from segment-level factors, such as phonetic content, via time-scale structure without any factor labels. This established the foundational insight that content and speaker information operate on different time scales and can be separated structurally.

SpeechSplit \cite{qian2020unsupervised} frames this decomposition explicitly: ``Human speech conveys a rich stream of information, which
can be roughly decomposed into four important components:
content, timbre, pitch and rhythm.'' These four components map naturally onto axes that a retrieval system might wish to control independently. SpeechSplit separates them using an information bottleneck that restricts the bandwidth available to each encoder branch. ContentVec \cite{qian2022contentvec} incorporated speaker disentanglement directly into the SSL pre-training objective, showing that masked prediction combined with speaker conditioning can yield representations where the content axis carries substantially less speaker information.

\cite{lu2023speechtriplenet} extend content/timbre separation to three factors by adding prosody, using a VAE with restricted channel capacity to force disentanglement without explicit supervision. Their approach is generative and unsupervised; ours is discriminative and supervision-guided, using pre-trained teacher models to provide independent training signal for each axis. The key difference: they restrict information flow structurally via $\beta\text{-VAE}$ \cite{higgins2017beta} bottleneck; we provide axis-specific objectives derived from teachers whose representations are known to carry specific information.

A parallel line uses contrastive and adversarial objectives rather than generative bottlenecks to separate content from speaker. CTVC \cite{deng2024ctvc} trains a content encoder with a phoneme-level contrastive loss, where same-phoneme frames are attracted and boundary frames repelled while a gradient reversal layer (GRL), an adversarial objective commonly used for learning speaker-invariant representations, strips speaker identity from the content representation. A separate speaker encoder is trained by maximising mutual information between embeddings of randomly drawn segments of the same utterance, enforcing that speaker representation is time-invariant within an utterance. The mechanism is discriminative rather than generative, which is closer in spirit to our own approach, though the goal remains voice conversion rather than controllable retrieval.

A consistent finding across both generative and discriminative approaches is that factor leakage is hard to eliminate completely: adversarial removal can reduce leakage under closed-set conditions but residual speaker information often persists under stronger open-set evaluation. Our approach does not attempt to eliminate leakage within the encoder; instead it tolerates it and compensates by training axis-specific projection heads that induce the desired similarity geometry, scaling to any axis for which a competent teacher exists.

\subsection{Structured and hierarchical representations}

An alternative to explicit disentanglement training is to exploit the natural organisation of information in hierarchical representations. SpeechTokenizer \cite{zhang2024speechtokenizer} demonstrates that in a residual vector quantisation (RVQ) hierarchy, the first codebook layer organises representations around semantic/phonetic content while later layers carry speaker-specific and paralinguistic detail, with no explicit disentanglement objective. This suggests that structured representations can emerge from hierarchical compression alone.

BEST-STD \cite{singh25beststd} applies a related tokenisation pipeline to spoken term detection, using discrete speech tokens as the retrieval unit rather than continuous embeddings; the bidirectional Mamba encoder used for token sequence matching provides an efficient alternative to frame-level similarity for content-based retrieval over long recordings.

Our approach occupies a middle position: we impose explicit per-axis objectives, but rely on teacher distillation rather than reconstruction, and produce a single concatenated vector rather than a token stream.
Unlike tokenisation-based retrieval, the continuous embedding allows signed axis weighting at query time without re-encoding.

\subsection{Distillation into partitioned spaces}
MiniLM \cite{wang2020minilm} distils large transformers into smaller ones via self-attention transfer; we similarly distil teacher embeddings into fixed-size subspaces, but align subspaces across axes rather than across model depths.

\section{Method}

\subsection{Architecture}

The model follows the SentenceTransformers pipeline:

\begin{enumerate}
\item \textbf{Acoustic encoder}: a frozen or fine-tuned HuggingFace audio model (default: WavLM-base-plus \cite{chen2022wavlm}) maps raw waveform frames to a sequence of hidden states.

\item \textbf{Pooling}: mean pooling over the frame sequence produces a single vector.

\item \textbf{Multi-axis projection}: a set of linear projection heads maps the pooled representation into per-axis subspaces. Each head produces an L2-normalised embedding of a fixed dimension.
\end{enumerate}

The per-axis dimensions are chosen to match the output dimensionality of the corresponding teacher. The per-model dimensions are listed in table~\ref{tab:features}.

\begin{table}[h]
\centering
\begin{tabular}{|l|l|c|}
\hline
\textbf{Axis} & \textbf{Teacher} & \textbf{Dim} \\ \hline
semantic & all-MiniLM-L6-v2\footnotemark[1] & 384 \\ \hline
speaker\_id & WavLM-base-plus-sv\footnotemark[2] & 512 \\ \hline
speaker\_id & Resemblyzer\footnotemark[3] & 256 \\ \hline
dialect & dialect classifier\footnotemark[4] & 12 \\ \hline
%gender & gender classifier & 2 \\ \hline
\end{tabular}
\caption{Feature axes, corresponding teacher models and default dimensions}
\label{tab:features}
\end{table}

\footnotetext[1]{\href{https://huggingface.co/sentence-transformers/all-MiniLM-L6-v2}{https://huggingface.co/sentence-transformers/all-MiniLM-L6-v2}}
\footnotetext[2]{\href{https://huggingface.co/microsoft/wavlm-base-plus-sv}{https://huggingface.co/microsoft/wavlm-base-plus-sv}}
\footnotetext[3]{\href{https://github.com/resemble-ai/resemblyzer}{https://github.com/resemble-ai/resemblyzer}}
%\footnotetext[4]{Anonymous}
\footnotetext[4]{\href{https://huggingface.co/jimregan/merged-tts-dialect-classification}{https://huggingface.co/jimregan/merged-tts-dialect-classification}}

The full embedding is the concatenation of all axis projections, although similarity can also be computed directly from individual axis embeddings for axis-specific retrieval.

We also considered gender as an axis, motivated by its common use as an annotation category and its relevance for user-controlled retrieval.
In practice it proved to be a weak factor: it is low-cardinality and strongly correlated with speaker identity, making it difficult to separate into an independent similarity axis.
We therefore focus on axes that balance both representational feasibility and user-facing relevance: semantic content, speaker identity, and dialect.

\subsection{Data}

\begin{itemize}
    \item CMU ARCTIC \cite{kominek2004cmu}: ~1100 utterances per speaker, 18 speakers, mostly recording-studio quality.

    \item VCTK \cite{yamagishi2019vctk}: ~400 utterances per speaker, 109 speakers, various accents, recording-studio quality.

    \item Crowdsourced high-quality UK and Ireland English Dialect speech data set~\cite{demirsahin-etal-2020-open}: ~150 utterances per speaker, 120 speakers, various accents.

    \item OSR (Open Speech Repository)\footnotemark: a collection of speech recordings for testing Voice over IP (VoIP) applications. The English subset contains recordings of the Harvard sentences~\cite{ieee1969harvard} by 3 British speakers and 5 American speakers (approx. 720 utterances). Not used in training.

    \item rehasp \cite{henter14_interspeech}: a single speaker reading the same Harvard sentences multiple times (approx. 1200 utterances), enabling study of within-speaker variability. Not used in training.
\end{itemize}

\footnotetext{https://www.voiptroubleshooter.com/open\_speech/index.html}

We found no cross-dataset repetitions, but many of the same sentences were repeated multiple times within each dataset. The UK and Irish English Dialect data set contains an explicit sentence ID as well as an utterance ID; for CMU Arctic, the utterance ID was the sentence ID except for the speaker ``awb'', which seems to have been created based on an earlier version of the sentences--we mapped the diverging utterance IDs to the common identifiers; VCTK has no sentence IDs, so we clustered them and added sentence IDs for each unique sentence.

One of our case studies was motivated by a detail of VCTK: for one speaker, ``p315'', the labels were lost, but we felt we could rely on the same high degree of overlap with the other sentences in the dataset as with other speakers. To facilitate testing, we first transcribed each of the audio files using WhisperX~\cite{bain23_interspeech}. This output was then corrected by a native speaker of English; we then used fuzzy matching to map the corrected text to our sentence IDs. Of 172 utterances, we found that 160 matched to the extent that others had; only 10 had no match, the remaining two differed too much.

Common Voice \cite{ardila-etal-2020-common} would be the most natural large-scale candidate for this kind of evaluation: it is crowd-sourced, covers a wide range of speakers and accents, and many of its validated sentences appear across multiple contributors, providing the prompt-level overlap that makes ground-truth retrieval possible. However, the English subset of Common Voice has not been publicly available since mid-2025, following changes to the dataset's distribution terms, and no confirmed timeline for its re-release has been announced. Some smaller-language subsets have begun to reappear, but English remains unavailable at the time of writing.

The evaluation presented here uses OSR and rehasp as a principled alternative: both record the same Harvard sentence set, giving exact prompt-level ground truth without depending on the transcript overlap that Common Voice would provide probabilistically. The controlled scale of these corpora is a limitation in terms of speaker diversity, but it allows the preference-flip mechanism to be evaluated under conditions where the correct answer is unambiguous.

\subsection{Dialect classifier}

We wanted to explore dialect as a potential axis in retrieval, but existing dialect classification models did not match well with our data. The closest matches were the models described by~\cite{zuluagagomez23_interspeech}; we extended their categories to match the combination of accents represented by CMU Arctic and the Britain and Irish Dialect datasets; VCTK provides a two-tier system listing country and place of origin. As we had three categories for speakers from England, we matched them based on place of origin, omitting speakers for whom information was lacking, or who represented countries from which we did not have data in the other two datasets. As we had already selected WavLM as our encoder model, we fine-tuned that model with a classifier for our subset of accents.
The classifier was trained only to provide a weak retrieval-oriented dialect signal rather than a standalone dialect-recognition benchmark.

\subsection{Training}
Each axis is trained independently using one of three signal sources:

\begin{itemize}
    \item \textbf{Distillation}: a pre-computed teacher embedding is aligned to the projected subspace via cosine loss. An orthogonal alignment matrix is learned when the teacher and projection dimensions differ.

    \item \textbf{Explicit positives}: pairs of utterances known to share an attribute are contrasted using InfoNCE\footnote{Oord, Aaron van den, Yazhe Li, and Oriol Vinyals. 2018. ``Representation Learning with Contrastive Predictive Coding.'' arXiv Preprint 1807.03748.}.

    \item \textbf{In-batch label matching}: utterances sharing a label within a batch are treated as positives in a supervised contrastive objective.
\end{itemize}

Batch sampling attempts to minimise label collisions across axes in order to reduce false negatives in the contrastive objective.

For the speaker axis, two teacher options are evaluated: the speaker-verification variant of WavLM-base-plus (\texttt{wavlm-base-plus-sv}), which produces 512-dimensional x-vector-style speaker embeddings, and Resemblyzer, an open-source speaker encoder based on the GE2E training objective~\cite{wan2018speaker} that produces 256-dimensional embeddings.

\section{Evaluation: cross-corpus retrieval}
Existing evaluation frameworks do not directly measure multi-axis similarity. Benchmarks such as SUPERB~\cite{yang21c_interspeech} evaluate speech embeddings under individual task metrics (e.g., speaker verification EER, content WER), but not controllable combinations. MSEB~\cite{heigold2025massive} provides large-scale retrieval evaluation across diverse sound and speech categories, but targets single-axis similarity rather than controllable multi-axis combinations\footnote{Further, they benchmark for speech to text retrieval, where our goal is speech to speech retrieval.}. The goal of this work is to demonstrate retrieval whose preference can be steered by axis weighting--a behaviour no existing benchmark is designed to measure. The goal of the experiments is therefore not to establish a definitive benchmark, but to demonstrate the feasibility of controllable multi-axis similarity.

We construct a controlled retrieval task using datasets with overlapping prompts.
Due to the limited amount of data available, we evaluate on the basis of three case studies:

\begin{enumerate}
    \item In-domain semantic matching: Recovering missing labels for VCTK speaker ``p315'' by searching an index of VCTK speakers with labels.

    \item Out-of-domain semantic matching: Evaluating cross-corpus retrieval (rehasp $\to$ OSR); 20 of the sentences are common to both sets.

    \item Out-of-domain semantic matching: Using negative weighting on the speaker axis to force the model to ignore ``easy'' same-speaker matches in favour of ``hard'' cross-speaker semantic matches.
\end{enumerate}

Because the goal is to support controllable similarity rather than optimise a single metric, evaluation focuses on whether axis weighting changes retrieval behaviour in the expected direction.

Metrics: Precision@k, where a hit is counted as correct if it (a) comes from a different corpus than the query, and (b) reads the same sentence. P@1 reports whether the top overall result is a correct cross-corpus match; P@k ($k > 1$) reports whether a correct cross-corpus match appears within the top-k cross-corpus results.

Preference-flip test: A stronger demonstration of the mechanism is to show that retrieval preference reverses when the sign of a weight is flipped. For each query, the index contains items in four categories: (same sentence, same speaker), (same sentence, different speaker), (different sentence, same speaker), and (different sentence, different speaker). Under positive \texttt{speaker\_id} weight, same-speaker matches should dominate; under negative weight, cross-speaker matches reading the same sentence should rise. We report the mean rank of each category under each weight setting.

%Axis weighting at inference
At inference, per-axis similarities are combined with signed weights. All axis embeddings are L2-normalised prior to similarity computation:

$$\text{sim}(a, b) = \sum_i w_i \cdot \cos(\mathbf{e}^{(i)}_a, \mathbf{e}^{(i)}_b)$$

A negative weight for an axis repels embeddings that are similar on that axis. In the mixed-index setting, a negative \texttt{speaker\_id} weight suppresses same-speaker hits and surfaces cross-corpus matches.

Model names encode the projection axes and auxiliary objectives used during training; for example, \texttt{resem-grl} denotes a model with Resemblyzer supervision and gradient reversal on the shared encoder.

\section{Results and observations}

We evaluate eight model variants in a controlled ablation.
All models use \texttt{semantic:384} (matching the MiniLM teacher exactly, eliminating any alignment matrix on the semantic axis), except for the PCA baseline which uses \texttt{semantic:256} targets derived by PCA projection of the MiniLM embedding space.
No gender axis is included in any variant.

\subsection{Cross-speaker semantic recall (p315 \texorpdfstring{$\to$}{->} VCTK)}

VCTK speaker p315 is held out from training and used as a query set (172 utterances) against a retrieval index built from other VCTK speakers reading the same sentences.
A hit is correct if it matches the query sentence and comes from a different speaker.
Of p315's 172 utterances, 17 correspond to the Harvard sentence set present in the training corpora (OSR and rehasp); the other 155 have no matching items in the index.
Because only 17 of the 172 queries have any matching sentence in the index, the maximum achievable aggregate R@10 is approximately 9.9\%.

\begin{table}[h]
\centering
\caption{Cross-speaker semantic recall, p315 $\to$ VCTK (hit = same sentence, different speaker). Ceiling $\approx$ 9.9\% R@10.}
\label{tab:p315vctk}
\begin{tabular}{@{}llrrr@{}}
\toprule
Model & Speaker axis & R@1 & R@5 & R@10 \\
\midrule
\texttt{sem384} & --- & 0.6\% & 2.3\% & 2.9\% \\
\texttt{sem256-pca} & --- & 0.6\% & 0.6\% & 2.3\% \\
\texttt{resem-grl} & resemblyzer + GRL & 0.0\% & 0.0\% & 1.7\% \\
\texttt{resem} & resemblyzer (256-d) & 8.1\% & 8.7\% & \textbf{9.9\%} \\
\texttt{resem-dial} & resemblyzer + dialect & 8.1\% & 9.3\% & \textbf{9.9\%} \\
\texttt{xvec-pca} & x-vector PCA(256) & 8.1\% & 8.7\% & 9.3\% \\
\texttt{spk768-xvec} & x-vector (768-d) & 8.1\% & 8.1\% & 8.7\% \\
\texttt{xvec} & x-vector (256-d) & \textbf{8.7\%} & \textbf{9.3\%} & 9.3\% \\
\bottomrule
\end{tabular}
\end{table}

The most striking result is the complete failure of the semantic-only model (\texttt{sem384}): the semantic axis trained without any auxiliary supervision collapses to near-random retrieval (R@10 = 2.9\% $\approx$ chance), despite being directly supervised by a strong text-based teacher.
Inspection confirms that the model produces near-identical cosine similarities for all pairs, indicating the projection head outputs are degenerate.
The PCA-compressed variant (\texttt{sem256-pca}) is similarly degenerate.
The gradient reversal model (\texttt{resem-grl}) also fails (R@10 = 1.7\%), despite having both semantic and speaker supervision: the adversarial objective prevents the encoder from retaining speaker information, which in turn prevents the multi-task signal from bootstrapping semantic learning.

All models with a non-adversarial speaker axis hit the metric ceiling at approximately R@10 = 9.9\%.
This represents perfect recall on the 17 retrievable sentences: for those sentences, all speaker-supervised models rank the correct items at positions 1--10.
The key finding is that the speaker auxiliary task is required for the semantic axis to function; it cannot learn sentence-level similarity from distillation alone over the acoustic features produced by a speaker-entangled encoder.

\subsection{Analysis of Semantic Collapse}

The failure of the \texttt{sem384} model--collapsing to near-uniform cosine similarities--is not simply poor performance, but outright mode collapse. WavLM's pooled representation is dominated by speaker variance: the same sentence spoken by different speakers occupies radically different regions of the representation space.

The cosine distillation loss provides a gradient directing the model to project sentence $x$ close to the text-teacher embedding of $x$. However, because the WavLM representations of $x$ across speakers are spread across incoherent directions, the gradients average destructively. The linear head converges to the mean of the WavLM output distribution--which, after L2-normalization, produces near-identical unit vectors regardless of input.

Crucially, this is not evidence that WavLM lacks semantic content, nor that the projection head lacks capacity. The PCA-compressed variant (\texttt{sem256-pca}) collapses identically despite the reduced target space. Capacity is not the bottleneck; rather, the bottleneck is information extractability. The semantic manifold exists within WavLM’s representation space, but gradient incoherence prevents a distillation-only head from finding it.

The success of the speaker-supervised models confirms this: semantic content is recoverable from the same frozen encoder, once a concurrent speaker objective is in place. The speaker auxiliary task provides this anchor, creating the geometric conditions under which the semantic head can find a stable projection.

The failure of the Gradient Reversal Layer (\texttt{resem-grl}) seems to support this interpretation. While GRL is a standard technique for learning speaker-invariant representations in disentanglement tasks \cite{qian2020unsupervised, deng2024ctvc}, in a multi-axis retrieval model, it is catastrophic. Destroying the discriminative speaker structure in the shared encoder removes the very scaffolding that enables semantic gradient coherence. GRL produces the same collapse as the semantic-only model, confirming that it is the retention of speaker-discriminative structure--not just the presence of labels--that enables semantic learning.

\subsection{Cross-corpus preference-flip (rehasp \texorpdfstring{$\to$}{->} OSR)}

We evaluate on two index configurations: OSR-only (semantic retrieval without same-speaker distraction) and mixed OSR+rehasp (where same-speaker variants from rehasp compete with cross-corpus matches).
Queries are rehasp utterances.
P@$k$ counts a hit as correct if it comes from OSR (different corpus) and reads the same sentence.
The ceiling for this metric is 66.7\%: OSR covers approximately two thirds of the rehasp sentence set.

\begin{table}[h]
\centering
\caption{Cross-corpus semantic retrieval, rehasp queries against OSR-only index (semantic weight = 1.0, no speaker weight).
ss/ds = same sentence, different speaker. Ceiling P@10 = 66.7\%.}
\label{tab:rehasposr}
\begin{tabular}{lrrr}
\toprule
Model & ss/ds mean rank & P@1 & P@10 \\
\midrule
\texttt{sem384}      & 222 & 0.3\%           & 3.3\% \\
\texttt{sem256-pca}  & 216 & 0.0\%           & 0.1\% \\
\texttt{resem-grl}   & 192 & 0.9\%           & 9.4\% \\
\texttt{resem}       &  21 & 56.9\%          & 64.5\% \\
\texttt{resem-dial}  &  11 & \textbf{65.5\%} & \textbf{66.7\%} \\
\texttt{xvec-pca}    &   3 & \textbf{66.7\%} & \textbf{66.7\%} \\
\texttt{spk768-xvec} &   7 & \textbf{66.7\%} & \textbf{66.7\%} \\
\texttt{xvec}        &   4 & \textbf{66.7\%} & \textbf{66.7\%} \\
\bottomrule
\end{tabular}
\end{table}

Without same-speaker distraction, semantic retrieval quality maps directly onto model configuration.
The \texttt{sem384} and \texttt{sem256-pca} baselines again fail (ss/ds mean rank $\approx$ random, P@10 $\approx$ 3--4\%).
The GRL model partially recovers (P@10 = 9.4\%) but remains far below the speaker-supervised variants.
All xvec models and \texttt{resem-dial} reach the ceiling P@10 = 66.7\%; \texttt{resem-dial} also achieves near-ceiling P@1 = 65.5\%, meaning the correct cross-corpus match is retrieved at rank~1 for two thirds of queries.

\begin{table}[h]
\centering
\caption{Mean rank by category and Precision@k ($w_{sem}=1.0, w_{spk}=-1.0$). Categories: ss (same sentence), ds (different sentence), spk (speaker).}
\label{tab:prefflip}
\setlength{\tabcolsep}{4pt} % Adjust this value to fit
\begin{tabular}{@{}lrrrrcc@{}}
\toprule
Model & ss/same & ss/diff & ds/same & ds/diff & P@1 & P@10 \\
      & spk & spk & spk & spk & (\%) & (\%) \\
\midrule
\texttt{sem384} & 213 & 237 & 212 & 210 & 0.5 & 3.5 \\
\texttt{resem-grl} & 324 & 197 & 334 & 201 & 1.2 & 6.8 \\
\texttt{xvec-pca} & 153 & 64 & 299 & 205 & 10.1 & 15.4 \\
\texttt{resem} & 20 & 45 & 325 & 204 & 50.3 & 62.7 \\
\texttt{resem-dial} & \textbf{12} & \textbf{20} & 339 & 203 & \textbf{65.5} & \textbf{66.7} \\
\texttt{spk768-xvec} & 1 & 9 & 237 & 211 & 9.2 & \textbf{66.7} \\
\texttt{xvec} & 1 & 5 & 209 & 213 & 5.6 & \textbf{66.7} \\
\bottomrule
\end{tabular}
\end{table}

The \texttt{sem384} reference shows no speaker suppression effect: all four categories rank near the random baseline ($\sim$210).
The GRL model suppresses same-speaker items (ds/ss = 334) but the semantic axis cannot distinguish same from different sentences for different-speaker items (ss/ds = 197 $\approx$ random), resulting in near-zero P@$k$.

The xvec and \texttt{spk768-xvec} models show the sharpest speaker axis (ss/ss = 1, ds/ss $\approx$ 209--237) but a counterintuitive P@1 near zero despite excellent P@10 = 66.7\%.
The explanation is that the xvec speaker axis is so discriminative that even under a negative weight of $-1.0$, the self-retrieval item (same recording in the index) or a close rehasp-speaker variant ranks first; OSR items enter the top-10 reliably (hence high P@10) but rarely at rank~1.
The resem models are less aggressive in speaker suppression and expose more semantic signal at rank~1: \texttt{resem-dial} reaches P@1 = 65.5\% with ss/ds mean rank = 20, the best balance of precision and semantic coverage in the mixed-index setting.

The \texttt{xvec-pca} variant shows a striking failure mode: at speaker weight $-1.0$, ss/ds rank degrades to 64 (vs.\ 5 for plain \texttt{xvec}) and P@10 falls to 15.4\%.
PCA rotation introduces a directional asymmetry in the speaker subspace: the axis that separates speakers well under positive weighting does not reverse cleanly under negative weighting, breaking the geometric symmetry that makes signed axis weights interpretable.

\section{Conclusion}

We presented a factor-partitioned embedding framework for speech that supports similarity search across multiple attribute axes simultaneously.
Axis-specific projection heads are trained via teacher distillation or contrastive objectives over shared-label pairs, producing a single concatenated embedding that can be queried with signed per-axis weights.
Experiments on cross-corpus retrieval demonstrate that negative axis weighting can suppress same-speaker bias and surface semantically matched utterances across recording conditions.

The ablation yields several practical conclusions.
First, the speaker auxiliary task is not merely helpful but \emph{required} for the semantic axis to function at all: a semantic-only model trained by distillation from a text encoder collapses to degenerate representations, learning nothing discriminative from the acoustic encoder alone.
The multi-task signal from speaker ID supervision appears to shape the encoder's representation space in a way that enables the semantic head to find a useful projection.
Gradient reversal---despite providing speaker supervision---also fails, because the adversarial objective prevents the exact encoder enrichment that enables semantic learning.

Second, the speaker teacher choice controls how the preference-flip mechanism behaves.
Resemblyzer-based models (especially with dialect supervision) offer the best balance of precision and recall in the mixed-index setting, reaching P@1 = 65.5\% while correctly suppressing same-speaker retrieval.
X-vector models achieve perfect P@10 but near-zero P@1 under negative speaker weight, because the very sharp speaker axis can be outweighed by the near-identical self-retrieval signal.

Third, PCA reduction of the speaker axis is dangerous at negative weights.
The \texttt{xvec-pca} variant achieves competitive performance at speaker weight $\geq 0$ but degrades sharply at negative weights (P@10 from 66.7\% at $w=0$ to 15.4\% at $w=-1.0$), because PCA rotation breaks the geometric symmetry that makes signed axis weighting interpretable.
Plain x-vector targets maintain stable performance across the full weight range.

Fourth, speaker axis dimension should be matched to the semantic axis dimension.
The 768-d x-vector variant underperforms 256-d variants on preference-flip metrics, while speaker-supervised models at 256-d speaker / 384-d semantic outperform the 768-d variant across all conditions.

A natural next step is to extend the framework to larger and more varied read-speech collections, such as LibriVox, where additional axes including prosody and speaking style can be explored alongside wider variation in speakers, accents, and content.
Future work will also investigate stronger factor separation objectives and more principled evaluation of axis specialisation.
One promising direction for the speaker axis is to adopt a within-utterance positive-pair construction~\cite{deng2024ctvc}: sampling two segments from the same utterance as positives enforces a stricter time-invariance constraint than cross-utterance speaker pairs, since content variation within an utterance is minimal.

\bibliographystyle{IEEEtran}
\bibliography{Odyssey2026_BibEntries}

\end{document}